# Are Face and Object Recognition Independent? A Neurocomputational Modeling Exploration

Panqu Wang, Isabel Gauthier, Garrison Cottrell

## Abstract:

Are face and object recognition abilities independent? While it is commonly believed that they are, Gauthier et al. (2014) recently showed that these abilities become more correlated as experience with non-face categories increases. They argued that there is a single underlying visual ability, $v$, that is expressed in performance with both face and non-face categories as experience grows. Using the Cambridge Face Memory Test (CFMT) and the Vanderbilt Expertise Test (VET), they showed the shared variance between CFMT and VET performance increases monotonically as experience increases. Here we address why a shared resource across different visual domains does not lead to *competition* and to an inverse correlation in abilities? We explain this conundrum using our neurocomputational model of face and object processing ("The Model", TM, Cottrell & Hsiao (2011)). We model the domain general ability $v$ as the available computational resources (number of hidden units) in the mapping from input to label, and experience as the frequency of individual exemplars in an object category appearing during network training. Our results show that, as in the behavioral data, the correlation between subordinate level face and object recognition accuracy increases as experience grows. We suggest that different domains do not compete for resources because the relevant features are shared between faces and objects. The essential power of experience is to generate a "spreading transform" for faces (separating them in representational space) that generalizes to objects that must be individuated. Interestingly, when the task of the network is basic level categorization, no increase in the correlation between domains is observed. Hence, our model predicts that it is the *type* of experience that matters, and that the source of the correlation is in the Fusiform Face Area, rather than in cortical areas that subserve basic-level categorization. This result is consistent with our previous modeling elucidating why the Fusiform Face Area (FFA) is recruited for novel domains of expertise (Tong, Joyce, & Cottrell, 2008).

**Key Words**: Face Recognition, Object Recognition, Neural Network, Perceptual Expertise, Individual Differences

# Introduction

Understanding how visual object recognition is achieved in the human visual cortex has been an important goal in various disciplines, such as neuroscience, neurophysiology, psychology, and computer science. Among all object classes, due to their social importance, faces have been studied most extensively, especially since the Fusiform Face Area (FFA) was discovered (Sergent, 1992; Kanwisher, McDermott, & Chun, 1997). Some research suggests that the FFA is a domain-specific "module" processing only faces (Kanwisher et al., 1997; McCarthy et al., 1997; Grill-Spector, Knouf, & Kanwisher, 2004); however, the FFA responds to non-face object categories of expertise, including birds, cars (Gauthier, Skudlarski, Gore, & Anderson, 2000; McGugin, Van Gulick, Tamber-Rosenau, Ross, & Gauthier, 2014; Xu, 2005), chessboards (Bilalić, Langner, Ulrich, & Grodd, 2011), and even artificial objects when subjects are sufficiently trained in the lab (Gauthier, Tarr, Anderson, Skudlarski, & Gore, 1999). High-resolution fMRI (HR-fMRI) in the FFA and neurophysiology in macaque's brain reveal the existence of highly selective face areas within the FFA or its likely homologue in monkeys, but no reliable selectivity for non-face objects (Grill-Spector, Sayres, & Ress, 2006; Tsao, Freiwald, Tootell, & Livingstone, 2006). However, when behavioral expertise is taken into consideration, more recent work found a reliable correlation between behavioral car expertise and the response to cars in the FFA, which remains reliable even in the most face-selective voxels in this region (McGugin, Gatenby, Gore, & Gauthier, 2012a; McGugin, Newton, Gore, & Gauthier, in press). They suggest that experience individuating members of a category may be sufficient to create this activation.

A more novel approach to study the relationship between face and object recognition is that of individual differences in behavioral performance. With the development of the Cambridge Face Memory Test (CFMT; Duchaine & Nakayama, 2006), reliable individual differences in face recognition abilities have been characterized in the normal population. Using a classical twin study design, Wilmer et al. (2010) provided evidence that face recognition ability is highly heritable. These authors also reported that face recognition ability (CFMT scores) shared very little variance (6.7%) with a test of visual memory for abstract art (AAM). In other work, performance on the Cambridge Car Memory Test (CCMT) was found to share only 13.6% of the variance with the CFMT, even though the two tests are very similar in format (Dennett et al., 2011). These results suggested that the ability to recognize faces has very little to do with the ability to recognize non-face objects.

Gauthier et al. (2014) challenged this conclusion by gathering evidence for the following hypothesis: face and object recognition share a domain-general visual ability, $v$, for discriminating visually similar objects, and this ability will only be expressed in performance when an individual has sufficient experience, $E$, for a given category. In brief, $Performance_{cat} \propto v \cdot E_{cat}$, where the subscript denotes a particular object category. The authors assumed that for faces, $E$ is generally saturated and makes little contribution to performance (as on the CFMT for instance). For objects however, they expected $E$ to vary much more across individuals, and as a result performance should not be as good a measure of $v$. But since they conceived of $v$ as the ability that allows people to learn from experience with a category, they predicted that $v$ would be expressed most directly in performance with objects in those people with the most experience. To test this hypothesis, the authors collected three measures from 256 subjects: 1) performance on the CFMT; 2) performance with eight non-face categories on the Vanderbilt Expertise Test (VET; McGugin, Richler, Herzmann, Speegle, & Gauthier, 2012b); and 3) a self-rating of experience with faces and the eight VET object categories (O-EXP, from 1-9).

For the CFMT, subjects studied six target faces and finished an 18-trial learning phase. They were then tested with 30 3-alternative forced-choice (3AFC) test displays to determine which faces were among the studied faces. They then studied the target faces again and were tested over 24 test trials, where the stimuli were presented in Gaussian noise. For the VET, subjects studied six target exemplars and then performed 12 3AFC training trials with feedback. Finally, they studied the six exemplars again and performed 36 3AFC (without feedback). In these trials, new exemplars from the target categories were used to test whether their learning generalized to new objects within the category.

Subjects were divided into six groups based on their level of reported experience with all VET object categories. According to their hypothesis, if the common visual ability $v$ is expressed through experience, then their performance on the VET (O-PERF) should also be more correlated with their performance on the CFMT as experience ($E$) grows. As predicted, a regression analysis found that as experience grows, the shared variance between the CFMT and O-PERF increased monotonically from essentially 0 to 0.59 along the six groups (See Figure 2(a)). The result indicated that the correlation is indeed moderated by experience: when subjects had sufficient experience with non-face objects, if they were found to perform poorly (well) with faces, they were found to also perform poorly (well) with non-face objects. This result suggests that data showing no or little correlation between object and face performance results from not taking into account the subject's level of experience with the objects.

These results are consistent with a neurocomputational model of face processing ("The Model" (TM); Dailey & Cottrell (1999); Cottrell & Hsiao (2011)). TM has been used to explain how and why an area of visual expertise for faces (the Fusiform Face Area) could be recruited for other non-face object categories: The resources in the face network can be shared with other object processing, provided that this processing is at the subordinate (expertise) level task (Joyce & Cottrell, 2004; Tong et al., 2008).

The present implementation of TM is similar to the expert network described in Tong et al. (2008): (1) images are preprocessed by Gabor filters, modeling V1; (2) the Gabor representation is analyzed by Principal Components Analysis, which we consider to correspond to representations in the Occipital Face Area; and (3) a neural network with one hidden layer is trained to recognize individual faces. The model is then trained on object categories at the subordinate level. I.e., we assume that experience with a category leads to recognition at the subordinate level (e.g., white, brown, and portobello mushrooms).

Since this is an individual differences study, one network corresponds to one subject. We used individual behavioral data from Gauthier et al. (2014), including CFMT scores, VET scores, and VET category experience scores. Because Gauthier et al. (2014) found self-report of faces to be less reliable than that for objects, we simply assumed that all subjects have a very large amount of experience with faces, so that their CFMT score represents their domain general ability $v$. We therefore identify $v$ with the CFMT score, and map that score to the number of hidden units. We map the self-rated experience score $E$ to the number of appearances of individual items within a specific category during training. As described above, we first train the network on faces to simulate the ability expressed by the CFMT performance, and then train on three non-face object categories (butterflies, cars and leaves) to simulate the abilities tested by the VET. We show that the shared variance between the recognition accuracy on faces and the average recognition accuracy on non-face objects increases as experience with the non-face object categories increases, consistent with Gauthier et al.'s data.

In Gauthier et al., the correlation with VET scores did not obtain when they used data from a single category on the VET. Instead, they had to average over the experience with all VET categories, which we replicated here. However, when we increased the number of subjects (networks), we found correlations based on single categories.

Consequently, we predict that the correlation between scores on the CFMT and on the VET will be observed depending only on experience with a single category, if enough subjects are tested. This prediction of the model has yet to be tested.

Furthermore, we show that the effect of experience moderating the correlation between VET and CFMT scores is not observed in our model if it is only trained to make basic level categorizations; hence we predict that this effect is carried by the FFA. This suggests that CFMT scores should have the increasing correlation with VET scores not just based on mere experience with a category, but the *kind* of experience with a category, where members of the category are processed at the subtype level.

Finally, we run an analysis on the net input of hidden units in two networks with different levels of experience, and show the power of experience is to expand the representational space to a larger region, where each individual object is more separated. The experience moderation effect is a direct reflection of this power. This phenomenon is also consistent with previous research using TM that demonstrates why the FFA is recruited for other domains of expertise (Tong et al., 2008).

## Methods

### Architecture of TM

In general, TM is constructed using four layers that represent the human visual system from low level features to high level object categorizations (Figure 1). Given an input (retina level), we first pass the stimuli through a layer of classical Gabor filter banks, which represent the receptive fields of V1 complex cells (Daugman, 1985). The Gabor filters are composed of five spatial scales and eight orientations. In the second layer, the Gabor filter responses are processed using Principal Component Analysis (PCA) to reduce the dimensionality and perform efficient coding. The PCA layer models the information extraction process beyond primary visual cortex, up until lateral occipital region (LOC). We think of this layer as the structural description layer from the classic model of Bruce & Young (1986), i.e., the level where the representation is suitable for face recognition and facial expression analysis. Since PCA can be implemented using a Hebbian learning rule (Sanger, 1989), we consider this step to be biologically plausible. The next layer is the hidden layer in the neural network. We consider the number of hidden units as the available resources for the task. At this layer, features are learned through backpropagation that are useful for the task. For example, if the task is to discriminate different faces, this layer will learn face-related representations adaptively through learning, and we can assume this layer corresponds to the FFA. If the task is to classify basic-level object categories, the layer will learn basic-level related representations, modeling those in the LOC. The fourth layer is the output layer, which represents the categories of the different objects. This simulates the category cells in prefrontal cortex. At each layer of the preprocessing network, there is a normalization step before giving them to the next layer. Each image pixel value is z-scored independently across the image set, the Gabor filters are normalized to be a percentage of the total responses of the 8 orientations for each location, scale, and image, and each principal component value is z-scored across the data set.

**Dataset and Preprocessing**

We use four object categories in all of our experiments: butterflies, cars, faces and leaves. The three non-face object categories are three of the eight VET categories. The reason we chose these three domains is that there are readily available datasets for these VET categories that include subordinate level labels. We collected the images from four separate datasets: 1) faces: the NimStim Face Stimulus Set (Tottenham et al., 2009, has 646 images across 45 individuals); 2) butterflies: the Leeds Butterfly Dataset (J. Wang, Markert, & Everingham, 2009, has 832 images across 10 species); 3) cars: the Multi-View Car Dataset (Ozuysal, Lepetit, & Fua, 2009, has approximately 2000 images for 20 models); 4) leaves: the One-hundred Plant Species Leaves data Set (Mallah, Cope, & Orwell, 2013, has 1600 images for 100 categories). For every object category, we randomly chose 16 images from each of 10 randomly selected subordinate level categories to form the training set (12 images per individual) and test set (4 images per individual[1]). We first transform all images to grayscale and crop them to a uniform size of $64 \times 64$ pixels. We then process them through Gabor filter banks as defined in Lades et al. (1993), with 8 different orientations ranging from 0 to $\frac{7\pi}{8}$ and five spatial scales. To make the filter response values in the same range, we normalize them across orientations for each scale on a per-image basis, so there is a low-frequency to high-frequency representation of the image. We normalize them across orientations for each scale on a per-image basis, so there is a low-frequency to high-frequency representation of the image. We normalize the response this way because we hypothesize that the downstream cells perform similar normalizations as the retina, which performs contrast normalization. In addition, this representation equalizes the power across spatial frequencies, so none dominate the representation.

We sample the 40 Gabor filter responses in an $8 \times 8$ grid over the image, resulting in a 2560-dimensional vector to represent a single image. The PCA step removes the redundancy of this representation by decorrelating the filter responses and generates a lower dimensional vector for efficient further processing. We perform PCA separately on the five scales, keep the eight eigenvectors with the largest eigenvalues for each scale, and project all Gabor filter responses for each image onto the corresponding eigenvectors. The 40 projections are z-scored by dividing by the square root of the corresponding eigenvalue before presentation to the neural network.

---

[1] Note that for faces, "individual" refers to a particular person, for butterflies and leaves, a particular species, and for cars, a particular make and model.

As in previous work (Tong et al., 2008), the label we give to the hidden layers (LOC or FFA) depends on the level of categorization. We hypothesize that LOC performs basic level categorization and FFA is involved in fine-level discrimination. As we showed in previous work, this changes the representation at the hidden layer dramatically, in that hidden units in the LOC model clump categories into small regions of representational space, while the hidden units in the FFA model increase within-category distance, spreading members of a category out into different locations in representational space.

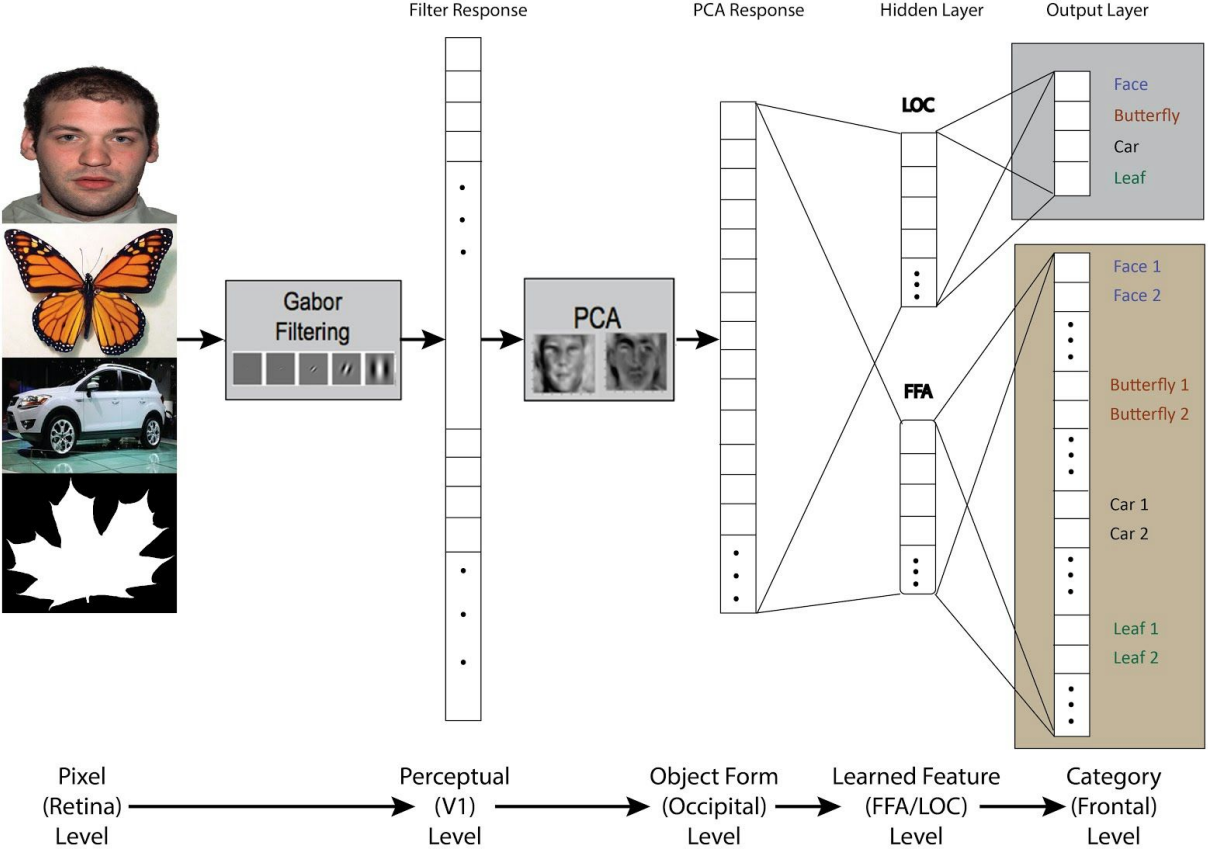

Figure 1. Model Architecture

**Mapping and Network Training**

To model Gauthier et al.'s experiment, we represent each subject by one network. The data for each of the 256 subjects are used to set the parameters for each network. In the psychological experiment performed by Gauthier et al. (2014), there are two key variables: the domain general visual ability, $v$, and the self-reported experience of the object categories, $E$. Based on Gauthier et al.'s theory, we write the following relation: $Performance_{cat} \propto v \cdot E_{cat}$. That is, $v$ is only expressed in performance via experience with a category. We can nominally think of $E$ as a number between 0 and 1, although this is transformed in our model to a more relevant range of experience for the networks. We assume that the maximum value of $E$ is the value for faces (every subject has maximal experience with faces), which means we can measure $v$ directly from each subject's data as their performance on the CFMT (a number from 0 to 1). $E$ is given directly in the self-report data (a number from 1 to 9).

We assume $v$ is based on the available representational resources of the subject for processing faces and objects, hence we map $v$ to the number of hidden units in each network using a simple function. With more hidden units, the network in general will generate higher dimensional and more accurate features for a given object category, thus improving the classification performance. We choose the particular mapping through cross-validation so that we do not use too many hidden units for the size of our dataset, which would result in poor performance from overfitting.

We use a linear function to map the reported experience (i.e., 3+$E$) to the frequency of individual exemplars in an object category in the training set. In Gauthier et al. (2014), the test-retest reliability for the self-reported experience measure, O-EXP, for non-face object categories is much higher than that of faces (0.60). As noted above, we imagine that face experience is maximal for each subject, and for the other categories, we use a linear mapping from the self-reported O-EXP, as the simplest possible unbiased estimate of the relationship between reported experience and training examples. Since in our database, we have 12 images each of 10 subordinate categories for each type (faces, cars, leaves, and butterflies), if a subject has experience level 1 with leaves, they will see 4 exemplars of each leaf, or 40 images of leaves. If they have experience level 9, they will see all 12 exemplars. We repeat the smaller number of exemplars to match the number of training instances in a model network's "day". Hence we are mapping O-Exp to the *variety* of experience with an object category.

For the faces, always use all 120 images of 10 people in the training set. The scaling above is calibrated to reach 480 updates of the weights per epoch, again, providing each network with an equal length "day." Hence, given a fixed training time (e.g., one epoch), different object categories have a different variety of training examples based on their level of experience. This mapping is reasonable given that more experience with a category should lead to more variety of experience with a category. Consider, for example, that a good chef will know many different varieties of mushroom, where a less-experienced cook may know only two or three.

As a result, our variable mapping and general training process of the network are as follows: we map $v$ to the number of hidden units, and $E$ to the amount of training examples that appear at each training iteration. For each network, we train on subordinate level face identification first, in order to simulate the process of gaining expertise on faces. This is intended to reflect the fact that before humans become familiar with the various species of butterflies, for example, they had expertise on faces. After training on individuating faces, we add the three non-face object classes (butterflies, cars and leaves) into the network by adding extra sets of output nodes and new training examples. In Experiment 1 and Experiment 2 in the next section, as the task is to discriminate the 10 individuals in each category, all networks have 40 output nodes. In Experiment 3, as we only perform basic level categorization of the non-face categories, so the network only has 13 output nodes, with 10 for individual faces and 3 for each non-face object category. We measure the recognition accuracy on the test set for each object when we finish training, and use this score to model the VET performance.

## Results

We will describe three simulations in this section. The first experiment is intended to model directly the psychological experiment performed by Gauthier et al. (2014), that showed that the correlation between performance on the VET and the CFMT increases with experience with objects. In that experiment, the level of experience was averaged across categories, because they did not find a correlation between performance on the VET for a single category based on experience with that category. The second experiment provides a prediction that, if more subjects were used, the correlation would emerge at the single category level. In the first and second experiments, the networks were trained to be "experts" in the categories, i.e., they were trained to individuate people, car models, and butterfly and leaf species. This suggests that the correlation emerges as a result of shared variance within the FFA. The third experiment predicts that we would not see the experience moderation effect based on basic level experience - expertise is necessary. Finally, we analyze networks trained to be experts to show *why* the experience moderation effect appears when using the same hidden units, counter to the intuition that there should be a competition for shared resources.

**Experiment 1: Modeling Gauthier et al. (2014)**
Gauthier et al. hypothesized a single underlying visual ability, *v*, that is only expressed through experience. This visual ability can be measured by performance on a face recognition test like the CFMT, as we all have a great deal of experience with faces. If *v* is a shared ability, it should become expressed in performance as a function of experience with non-face objects.

To model their experiment and results, we make an one-to-one mapping of $v$ and $E$ to our neural networks, with each network representing one human subject. Since $Performance_{cat} \propto v \cdot E_{cat}$ (according to Gauthier et al.'s hypothesis), and every human subject is assumed to have high and relatively similar experience with faces, their $v$ is explicitly expressed by their face recognition score on the CFMT. We therefore initialized the network based on the subject's CFMT score by mapping that number to the number of hidden units according to the following formula:

$$N_{hidden}(s_{net}) = \lfloor 34 \times CFMT(s_h) - 14 \rfloor$$

Where $s_h$ represents a particular human subject, $s_{net}$ is the corresponding network modeling that subject, $CFMT(s_h)$ is the percent correct of $s_h$ on the CFMT, and $N_{hidden}(s_{net})$ is the number of hidden units for that subject's network. The CFMT scores in Gauthier et al.'s data range from 0.4722 to 1, so $N_{hidden}$ ranges from 2 to 20. As in general $N_{hidden}$ must be matched to the size of the dataset for good generalization, our range of hidden units is chosen by cross-validation to ensure that there are sufficient resources at the maximum number to provide good generalization without overfitting.[2]

Similarly, the formula for mapping self-rated experience (O-EXP) to the number of training samples for each subordinate object category is as follows:

$$N_{sample}(category, s_{net}) = 3 + O\text{-}EXP(category, s_h)$$

---

[2] In general, the number of hidden units depends upon the size of training set. In recent winner of ImageNet Large Scale Visual Recognition Challenge, the networks are trained with over 1.2 million images, and the final hidden layer has 4096 units (Krizhevsky et al., 2012). However if the same network is trained on a smaller dataset, the recognition accuracy is low due to overfitting. (Zeiler & Fergus, 2014).

As O-EXP ranges from 1 to 9, then the number of training samples ranges from 4 to 12 (12 is the maximum number of individual training samples in the dataset for each individual). Hence, we use a fraction of the dataset to learn each object when the subject has lower experience, while we use the full dataset to train the networks with the highest experience. For faces, we assume O-EXP is 9. Note, as described above, we must ensure that the networks are trained with the same total amount of images per epoch so that every network has same number of updates. That is, there are the same number of "hours in the day" for each network. We set this number to 480, as this is the size of the most diverse training set (120 images of 10 individuals for 4 categories). We use $N_{sample}$ to compute a proportion of the dataset. I.e., assuming leaves and cars are the only two object categories for the moment, if $N_{sample}$ for leaves and cars is 6 and 12, respectively (with $N_{sample}=12$ by definition for faces), the proportion of the training set that are leaf images is 6/(6+12+12), or 20%.

We use stochastic gradient descent (online-backpropagation) to train the network. A learning method with equivalent results to backpropagation, contrastive hebbian learning (CHL), can be implemented in a biologically plausible way (O'Reilly, 1996; Plaut & Shallice, 1993). While less biologically plausible, backpropagation training is much more efficient than CHL. The input vectors are z-scored, and the weights are drawn uniformly from the range -0.5 to 0.5. In all experiments, we set the learning rate to 0.015 and momentum to 0.01. As mentioned in the Methods section, we train the network on individuating faces first. We stop the face network training in either one of two conditions: if it hits the stopping threshold (mean squared error of 0.005, determined using cross-validation to provide the best generalization), or if the number of training epochs reaches 100, when we assume the network has gained sufficient expertise on face recognition as the training time is enough. We then start the second training phase by introducing the three non-face object categories into the training set and add 30 output nodes, corresponding to subordinate-level categorization of the 10 individuals in the 3 categories. The network is trained until the error is below 0.005, or training epoch reaches 90. At the end of the training process, we measure the recognition accuracy on the test set for all four object categories, and calculate the correlation between the score on faces and averaged non-face objects. We show the result in Figure 2(b).

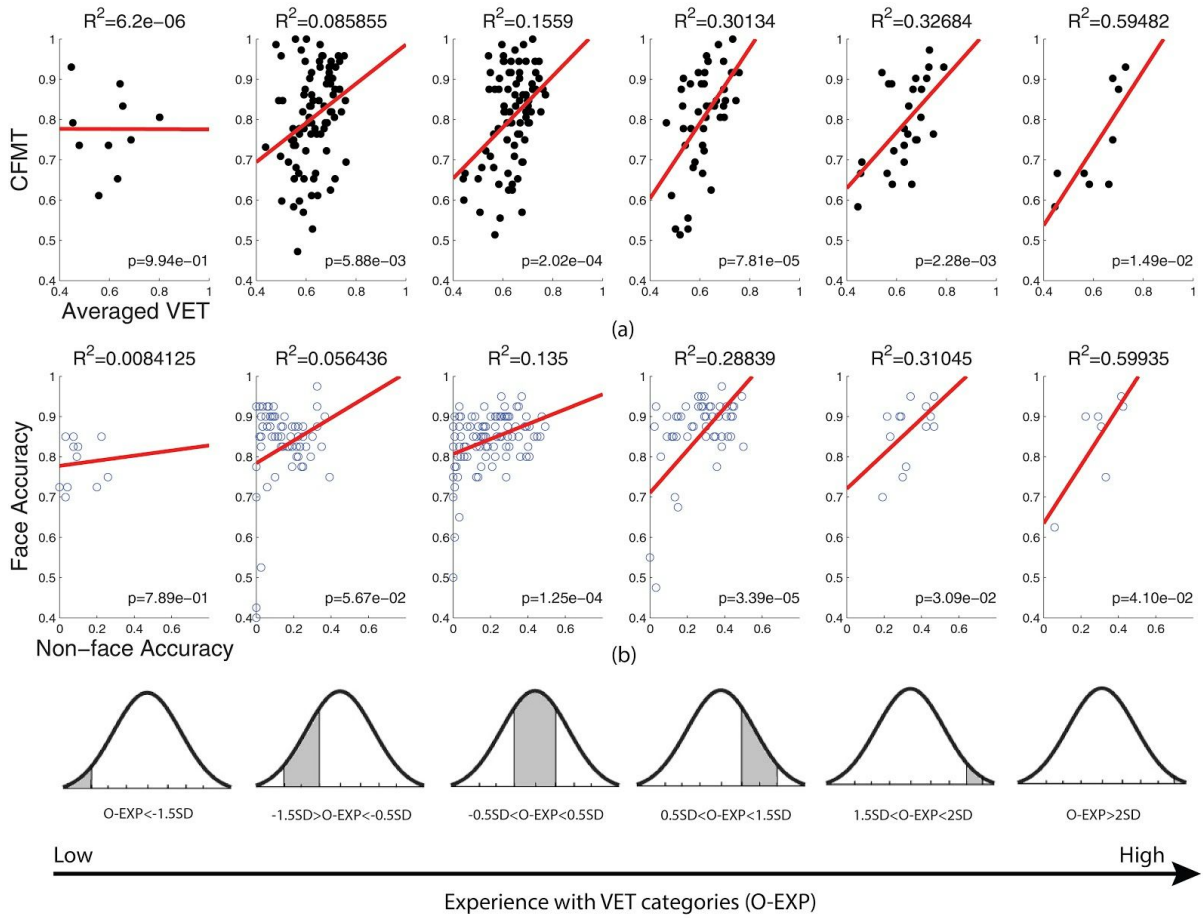

Figure 2. Results of Experiment 1. The first row (2a) shows the experimental data from Gauthier et al. (2014). The second row (2b) shows our modeling result. Each dot in 2b represents a single subject network whose parameters ($v$ and $E$) are calculated based on the corresponding human subject. Each line in the graph represents the regression for each group between their CFMT scores against their VET or non-face object recognition scores. The bottom row shows how the subjects are divided into six groups based on their self-rated experience score in VET object categories (O-EXP). For example, the second column (top row) shows the data from subjects (dots) whose O-EXP score is between 1.5 and 0.5 standard deviations below the mean.

From Figure 2(b), we can clearly see that as experience (O-EXP) grows, the shared variance between face recognition performance and the averaged non-face object recognition accuracy increases monotonically from $1.2 \times 10^{-4}$ to $0.60585$. This result matches those of Gauthier et al. qualitatively and demonstrates that our network training strategy and variable mapping of $v$ and $E$ are reasonable. The mapping of $v$ to various numbers of hidden units in the network spans the accuracy of face recognition (y axis of Figure 2(b)), suggesting that the hypothesis that the variance across individual subjects in the domain general object recognition ability is the amount of representational resources in cortex (hidden units in the neural network). The mapping of $E$ to the number of training examples on non-faces spans the accuracy of non-face object recognition (x axis of Figure 2(b)), clearly illustrating that higher experience will generally facilitate object recognition performance by moving them from all being relatively low to a range of scores, expressing the underlying computational resources.

**Experiment 2: Correlation with a Single Category**

In Gauthier et al. (2014), the increasing trend of correlation was not observed for any individual category. Rather, it only appeared for the *averaged* VET score (O-PERF) against the CFMT score. This is theoretically problematic because, according to their hypothesis, $v$ is a domain-general visual ability, and face recognition should not be independent of any non-face object category when people have sufficient experience in that category. In the original study, this situation was attributed to the fact that self-reports were likely very imperfect measures of experience with a category. However, in the present simulations, experience had a very direct mapping to each network's training and yet we also did not see the phenomenon as clearly in our simulations when using individual categories (see Figure 3). One possible explanation is that more subjects are required to show the effect; as there are few "experts" in the general population. In this experiment, we use a much larger number of subject networks and ability levels. We expect to see the same experience moderation effect as in the averaged category result if our assumption is true.

In this experiment, we use 1,000 different networks rather than the 256 in the previous experiment. To produce a larger range of network performance, we extended the range of hidden unit numbers and experience levels. We manually created the initialization of the values of $v$ and $E$ for the subject networks. We map $v$ to the range $N_{hidd} \in \{1,3,5,7,9,12,15,18,21,24,28,32,36\}$. We determined in advance that there is still no overfitting with up to 36 hidden units. For $E$, we set the range of experiential variety to $N_{sample} \in \{2,4,6,8,10,12\}$. As before, higher numbers of samples indicates more varied experience with that category. The number of subject networks at each level of $E$ and $v$ are determined by a Gaussian distribution, and the number of training examples falls in the given interval from 2 to 12. This approach tends to assign more members to the middle value in the set, simulating the fact that most people should have intermediate level of $E$ and $v$. The training procedure, dataset we use, and network parameter settings are the same as in Experiment 1. We show our result in Figure 4.

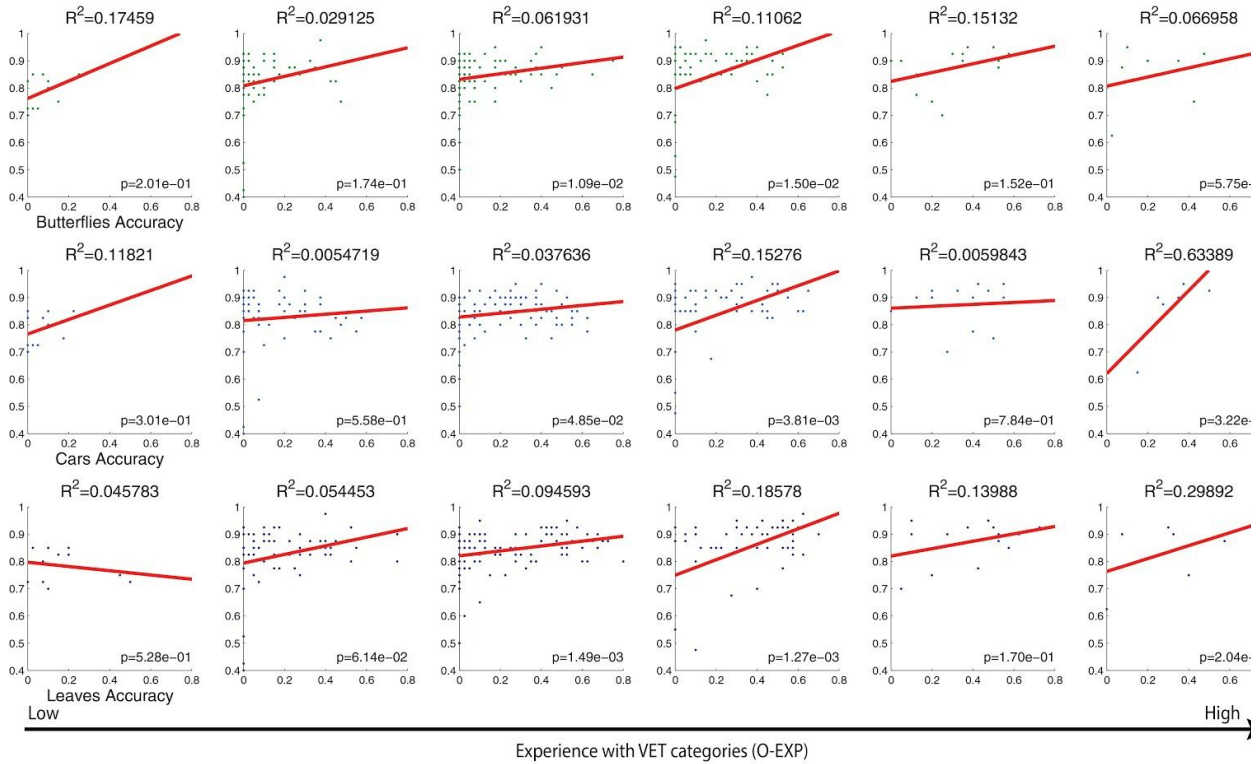

Figure 3. Result showing the correlation between the networks' face recognition performance and *single* non-face object recognition performance (butterflies, cars and leaves) in Experiment 1, as a function of experience. Interestingly, while there appears to be an overall trend of increasing correlation (especially for the leaves), it is generally smaller, and not monotonic when compared to the result using averaged performance (Figure 2(b)).

As can be seen from Figure 4, as experience grows, the shared variance ($R^2$) between face and all three individual non-face objects increases monotonically, from a value near zero ($p > 0.1$) up to a value greater than 0.7 ($p < 5 \times 10^{-5}$). Not surprisingly, when we calculate $R^2$ between face and averaged non-face performance, the increasing correlation trend still exists, from 0.048 ($p=0.1873$) to 0.829 ($p < 10^{-6}$). We ran the experiment 10 times, and the increasing correlation trend is very robust. The number of subjects is one factor in observing the experience moderation effect at the single-category level. A possible explanation for this finding is that using the averaged category experience leads to an aggregation effect (Rushton et al., 1983). At the single-category level, the smaller amount of data at any level of experience will be more variable, due to factors such as different initial random weights, different local minima, noise, etc. With several categories, these uncorrelated sources of noise are reduced. With more subjects at any given level of experience, we can also eliminate this nuisance variance, as long as it is not correlated across different subjects with similar experience, in the same way as it was not correlated across different categories for the same subjects. Our finding predicts that if more subjects were recruited, the experience moderation effect would be found at the single category level in actual behavioral data.

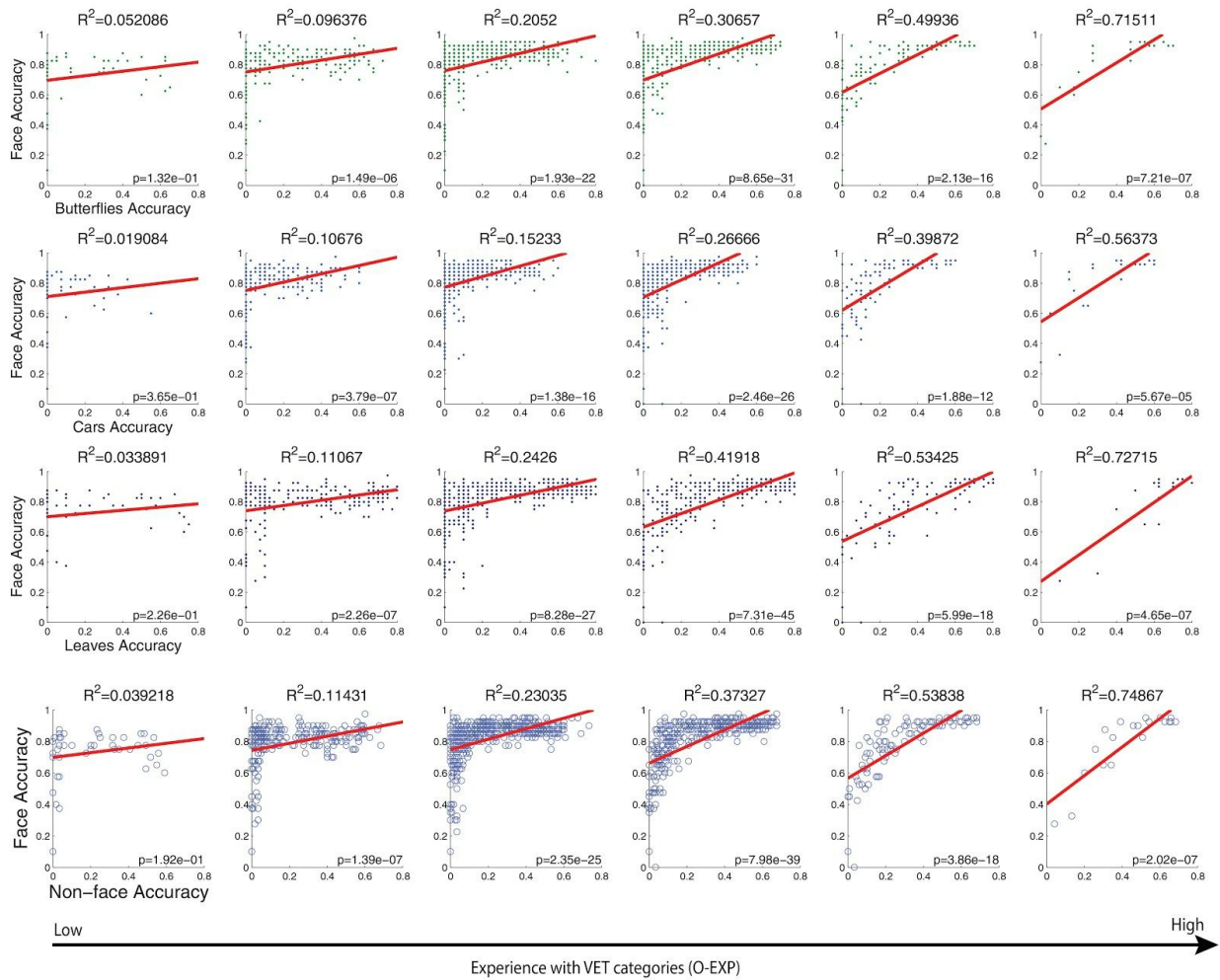

Figure 4. Results of Experiment 2. The top three rows show the trend of shared variance between face recognition accuracy (y axis) and *single* non-face object recognition performance (x axis, butterflies, cars and leaves for each row), as a function of experience. The last row shows the correlation on *averaged* non-face object recognition performance. Each dot represents a subject network, and the red regression curve is also plotted for each group. As we can see, the correlation is monotonically

increasing when experience grows, regardless of whether individual or averaged performance are used.

**Experiment 3: Basic-Level Classification**

In Experiments 1 and 2, the networks representing each subject are trained and tested on subordinate-level classification tasks, which means their job is to discriminate individuals (Is it Tom/John/Mary… or Benz/Ford/Toyota...?) within a category (faces or cars). That is, the networks are trained to become experts in these specific tasks. Based on our previous modeling (Tong et al., 2008) and fMRI (Wong, Palmeri, Rogers, Gore, & Gauthier, 2009b) work, we would expect the Fusiform Face Area to be a main site for such subordinate level learning. This however begs the question, is the overlap in abilities we and Gauthier et al. (2014) have measured depend on expertise at the subordinate level? In other words, would we see the same result of experience moderating the relationship between face and object recognition if the networks were instead trained on basic-level categorization?

Hence, in Experiment 3, we test this hypothesis by performing the same experiment on our networks, but training the network that has been pre-trained on faces to classify the objects at the basic level. In a previous modeling study (Tong et al., 2008), they analyzed the effect of both subordinate-level and basic-level classification tasks using the same neurocomputational modeling approach we use here, and found that there is a large difference in the hidden layer representational space developed over training in basic versus expert-level categorization. Here, we investigate the result of an expert network (a face identification network) being additionally trained to be a basic level categorizer, and we compute the correlation between face identification performance and basic-level categorization performance within the same network. If we still observe the experience moderation effect, it would indicate that the experience moderation effect is not specific to the subordinate level; if not, it suggests that the experience moderation effect requires that subjects' experience be at the level of subordinate-level categorization, or at least rules out that it works for just any training task.

To model the basic-level classification task, the only change we make from Experiment 2 is altering the number of output nodes and collapsing across individuals. We keep training the output nodes for faces to make sure the model remains effective at individuating faces. As we have 10 individuals for each of the 4 object categories, all networks in Experiment 2 have 40 output nodes; here, the networks only have 13 output nodes (10 faces + 3 nodes representing each non-face object category: butterflies, cars and leaves). The variable mapping and training procedure are otherwise exactly the same as Experiment 2. The result is shown in Figure 5.

As can be seen from Figure 5, as experience grows, we do not observe increasing correlation between face and non-face recognition performance, no matter whether experience is measured based on a single category or across categories. Instead, we observe a relatively constant correlation between performance in the two domains, regardless of how much experience the network has on objects. For the correlation results on single categories, we either find no correlation (leaves) or non-monotonically increasing low correlation (butterflies and cars). When performance is averaged across categories, however, due to the effect of aggregation, the overall correlation increases to around 0.35; nevertheless, the correlation does not monotonically increase as experience grows.

This phenomenon is easily explained. In Experiment 1 and 2, the variation across domain general visual ability ($v_v$) allows the networks to express the full range of face recognition ability, with the face recognition performance spread out between 0 and 1 (y axis in Figure 2, 3, and 4). However, due to the constraint of experience for the non-face objects, the network cannot express the full range of object recognition ability until the experience level is high. This can be seen from the results in Experiment 1 and 2 (x axis in Figure 2, 3, and 4), where the dots are "squeezing" around zero for low-experienced objects, and they gradually spread out when experience increases. In general, the cause for low recognition performance is either that the subject network has low $v_v$ (few hidden units), or because the subordinate-level task is very hard, and the resources are not sufficient.

In basic-level categorization, however, the task is easier (the networks only have to recognize all leaves as leaves, all butterflies as butterflies, etc.), and to do so the networks do not need a large number of hidden units, nor do we need very much training. Hence *all* of the networks (and by inference people) have enough resources to attain a relatively high score on basic level object recognition. This is shown clearly in Figure 5: face recognition performance is spread out as usual (y axis), and object recognition performance (x axis) has much lower variance in general. This explains why the correlation in the low-experience bins is approximately the same as in the high-experienced bins, and the increased in correlation with face recognition performance from the lowest level of experience (0.32) to highest level of experience (0.41) is not as large as in subordinate-level classification (Figure 4, from 0.05 to 0.83). Experience does not mediate performance in an easy task such as basic-level recognition, as the performance is dominated by the relative easiness of the task.

Hence we infer that the *type* of experience matters in deciding how abilities in different domains overlap: knowing the kind of leaf, or car, or butterfly leads to an increasing correlation of performance with face recognition, while just knowing that a leaf is a leaf, etc., does not. The level of task, even if both tasks involve categorizing images, has significantly different impacts on the outcome of the experiment. The need to differentiate between individual objects within a visually homogeneous category, rather than placing them into categories that differ in the overall part structure, produces the moderation effect shown in Experiment 1 and 2.

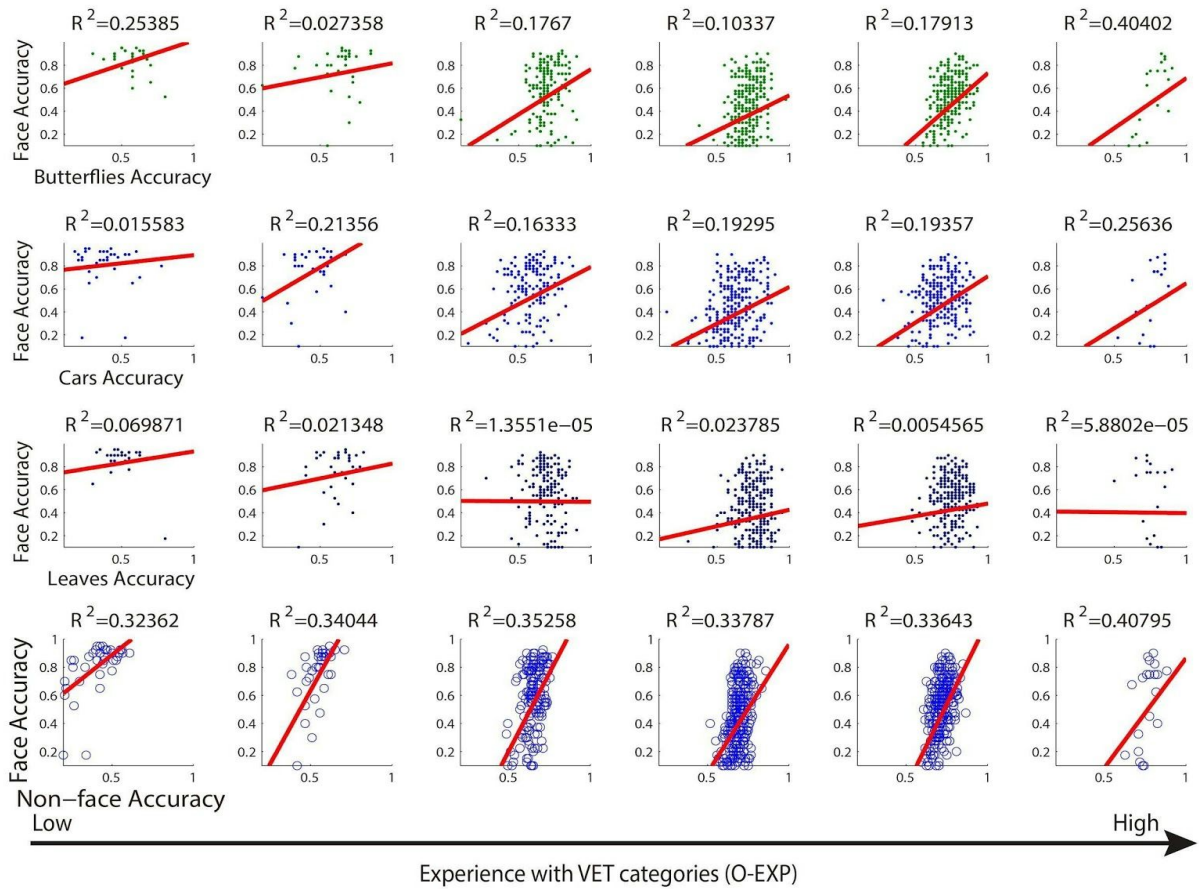

Figure 5. Result of Experiment 3 (basic-level classification). The format is the same as Figure 4, with top three rows shows the correlation between performance on face and single non-face objects, and the last row on averaged non-face objects. There is no monotonically increasing correlation in either single or averaged category performance.

**Analysis: The Power of Experience**

Given the finding that more experience leads to higher correlation between subordinate-level classification tasks in Experiment 1 and 2, we may wonder why this happens. For example, it seems intuitive that if the same hidden units are being used in both tasks, then there should be *competition* for these representational resources, and higher performance on faces should mean that more hidden units are dedicated to faces, which would result in lower performance on objects. This turns out not to be the case. Tong et al. (2008) showed that the hidden unit representation learned in a face identification task separates faces in hidden unit space, making it easy for a classifier to separate them. However, this same "spreading transform" *generalized* to novel categories. For example, they showed that when novel objects ("Greebles") were presented to the trained network for the first time, without any training, they were already spread out in hidden unit space. In this experiment, using a similar analysis of the net inputs of the hidden units, we show how this effect develops as a result of experience.

More specifically, we analyze the hidden units on two subject networks with different levels of experience. Recall that we map experience ($E$) to the number of training examples per individual. For this analysis, we set the number of training examples per object for the two networks be 3 and 12, respectively, representing low and high levels of experience. Both networks have 50 hidden units, so they have sufficient ability ($V$ $v$) to give the best performance. We train both networks on individuating faces first, and continue training on recognizing mixed object categories. We measure the performance at the end of training. During training, we record the net input of the hidden units for all training examples at six different time points (see Figure 6), which enables us to observe the evolution of the internal representation. For data collected from each subject network, we perform PCA on them and visualize the projections on the first and second principal components on a two-dimensional subspace. The result is shown in Figure 6. Note that for columns 1 and 2, the different colors represent different faces, to show how the faces are separated in the space. While some faces look like they are close in the space, they are separated by other dimensions. For columns 3-6, the different colors represent different categories.

Several conclusions can be drawn from the results. First of all, for the networks trained on face recognition only (the first two columns), no difference in experience exists, so the representations that the networks develop are similar: training on differentiating faces gradually separates each individual face in the subspace (second column), compared to the initial cluster at the center (first column). Second, when we take a close look at the third column, the non-face objects are already dispersed to the extent of the representational space formed by faces, even without training. This suggests that the projection into the hidden unit space learned for faces, which spreads out the faces, generalizes to novel objects, spreading them out as well. This is the same finding as in Tong et al. (2008), where it was shown that Greebles were spread out by the face network before it was even trained on Greebles. In that paper we also showed that there was nothing special about faces *per se*, rather, it is the task that is learned (individuation of similar looking items) that leads to this spreading transform. This result held for our model of the FFA, which suggests that the effects found in the Gauthier et al. (2014) paper are also a reflection of expertise with the non-face categories. Finally, when training on multiple object categories, we find that more training generally produces a larger spreading effect for both networks (the change from the third column to the last column), but more experience spreads the objects to an even greater extent (compare the last columns in the two rows). In data not shown, both of these networks achieve 87.5% accuracy on face recognition but the network with less experience with objects only achieves an average accuracy of 16.67% on non-face objects. This is well above chance, but much lower than the more experienced network, which achieves an accuracy of 83.33% on objects. As a result, we can speculate that greater experience actually leads to a greater spread in the hidden units of the network, and this spreading transform positively correlates with performance on the object recognition task. Performance on objects and faces is similar in a network with more experience, and very different in a network with less experience, as we saw in Figures 3 and 4. This is the power of experience.

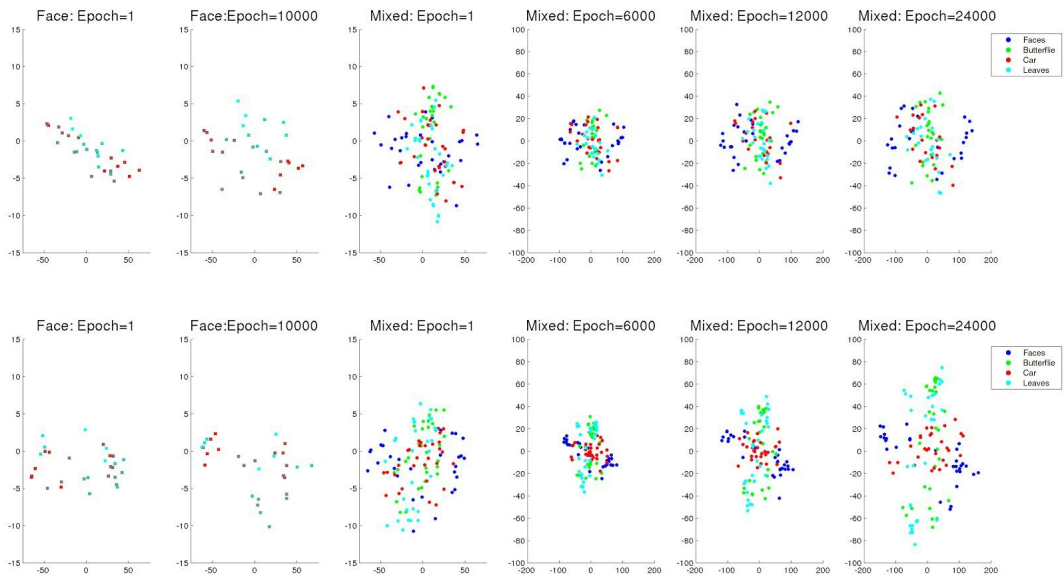

Figure 6. Visualization of the development of net input of hidden units over network training. First row: subject network with low experience (3 training examples per individual). Second row: subject network with high experience (12 training examples per individual). Each column represents the data collected from corresponding training epoch (shown in the title). In the left two columns, the colored dots represent different

individual faces. In the right four columns, the colored dots represent different object categories, shown in the legend. Note: The y-axis changes from [-15,+15] to [-100,+100] in the fourth column for clarity.

The above analysis is based on the PCA projection of the net input on a two-dimensional space. Since there are 50 hidden units in the network, we want to explore whether the phenomenon could generalize along all dimensions. As we cannot visualize a 50-dimensional space, we take five measurements for each dimension to help us understand its behavior:
1) Max: the maximum value of the projection on a principal component dimension, for each single category (locally) and across all of the categories (globally).
2) Min: the minimum value along a dimension, for each single category and across all of the categories.
3) Var: the variance along each dimension, for each single category and across all of the categories.
4) Inter: the average between-class distance, measured using Euclidean distance between the center of each object within the same category to the center of the current category, and averaged across all categories.
5) Intra: the average within-class distance, measured using Euclidean distance between each data point belonging to a single individual to the average of that individual's locations, and averaged across all categories.

Among the five measurements, Max, Min and Var are measured both globally (across all categories) and locally (for each object category), while Inter and Intra are only measured globally. Max, Min and Var indicate how far the individual representations are spread out along each dimension, while Inter and Intra measure the behavior for each group. The results are shown in Figure 7.

From the local measurement results in Figure 7, we can clearly see that:
1) For Max and Var, the value of high-experience network is always greater than low-experience network.
2) For Min, the value of high-experience network is always smaller than low-experience network.

These findings hold for all four object categories. These results demonstrate that for individual representations, high levels of experience separate them along *all* dimensions in the space.

For the global measurement (combined all categories), we can see that

1) For Max, Min, and Var, their behavior are the same as local measurements above.
2) For Inter and Intra, the value of high-experience network is mostly greater than low-experience network.

Imagine each object forms a cluster in the space. The Inter and Intra results indicate that as experience grows, each individual resident within that cluster will become further apart from its neighbors and the whole cluster itself will also move away from other clusters, like the "redshift" phenomenon in physical cosmology (Hubble, 1929). As this "redshift" of object representation happens in all dimensions of the hidden unit universe, it suggests that the essential power of experience is to generate a spreading transform for objects in the representational space, and accordingly to facilitate a subordinate-level classification task. The experience moderation effect, as can be seen in our experiment, is a direct outcome/reflection of this internal power, in a large population of subjects.

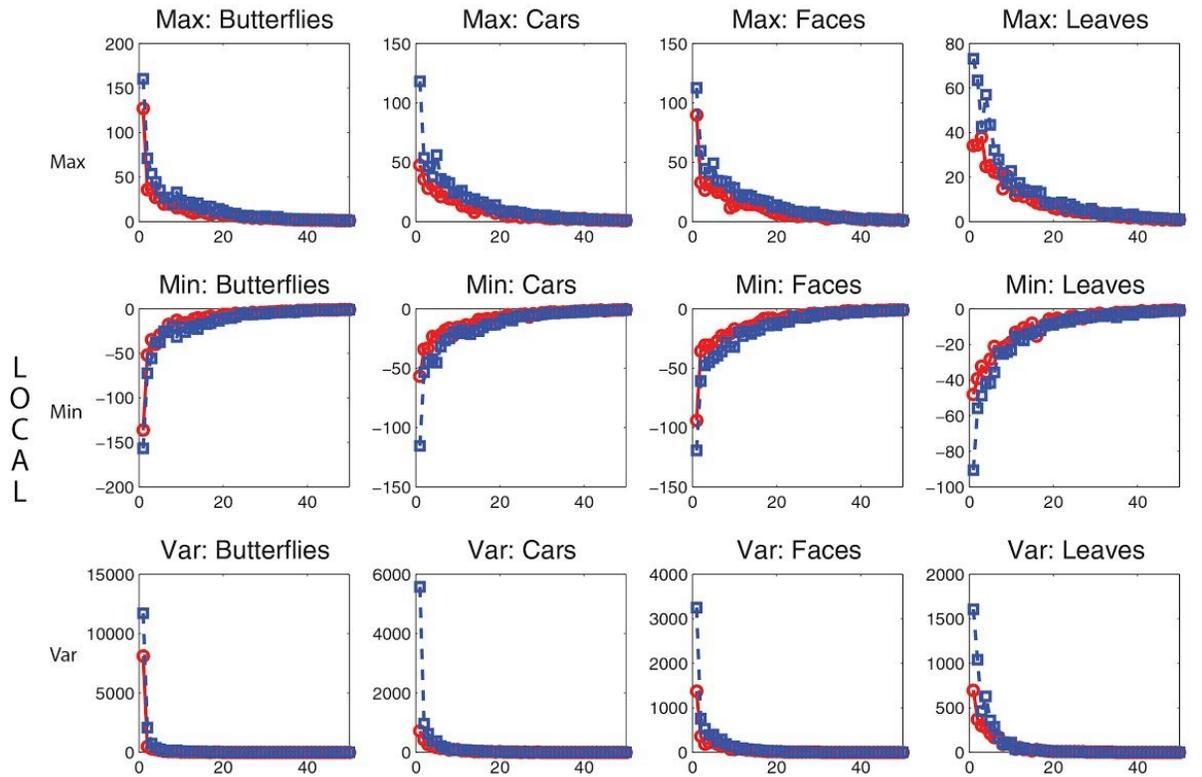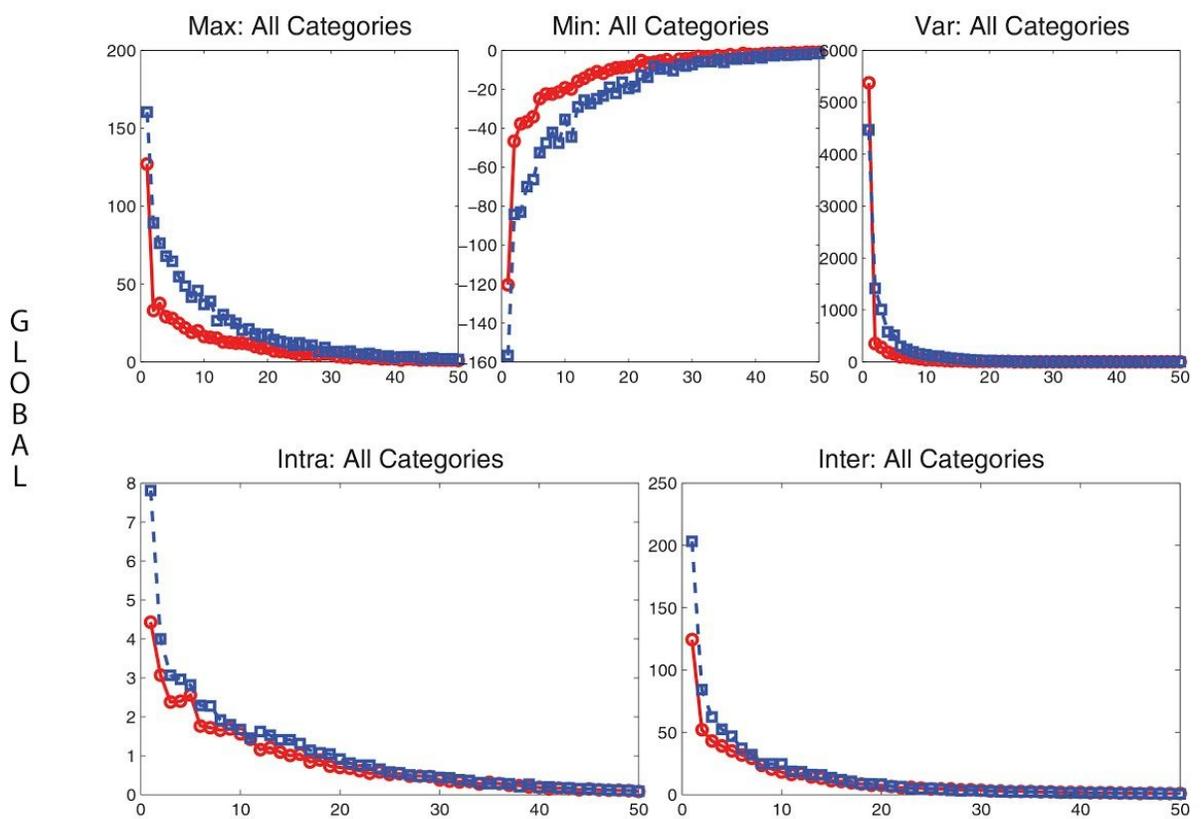

Figure 7. Visualization of the measurements taken on all 50 PCA dimensions of the hidden unit net inputs. In all graphs, the blue lines represent the network with a high level of experience, and the red lines represent the network with a low level of experience. We take five measurements: Max, Min, Variance, Inter-group distance and Intra-group distance, as described in the text. The top three rows show the result of Max, Min and Var on the four object categories (left to right: faces, butterflies, cars and leaves). The last two rows show the result of all measurements on all categories.

In addition, we measured the entropy of the net input of all the hidden units. Entropy is a measurement about how much information is preserved in the hidden units, and it is scale-free. If the data is highly scattered, the variance will be high, and more information will be carried. In order to calculate the entropy, we obtained the net input value of all hidden units across examples in the training set. We then calculated the entropy for each of the hidden units by getting the probability distribution (the normalized histogram) of the values, thereby computing $p_i$ for each bin, and then summing $p_i \log p_i$ over the bins (the results were robust across various bin sizes) We then averaged the entropy over all of the hidden units. To examine how the entropy develops over time, we plot its value as a function of training iterations, as shown in Figure 8. As we can see, although both networks show a general increasing trend of the entropy, the network with more varied experience always has higher entropy. This result is expected based on the PCA visualization in Figure 6, as the representations for both face and non-face objects become more separated as training proceeds. Again, this result demonstrates that the power of experience is to learn a more separated representation for objects to facilitate the subordinate-level classification task.

Furthermore, when looking into the local and global measurement of variance in Figure 7, we can see that for the more-experienced network, a larger number of dimensions accounts for more variance than for the less-experienced network. This suggests that the more-experienced network contains more complex information that must be decomposed into several different dimensions, which provides another way of measuring how the network is spreading out the representation of the categories.

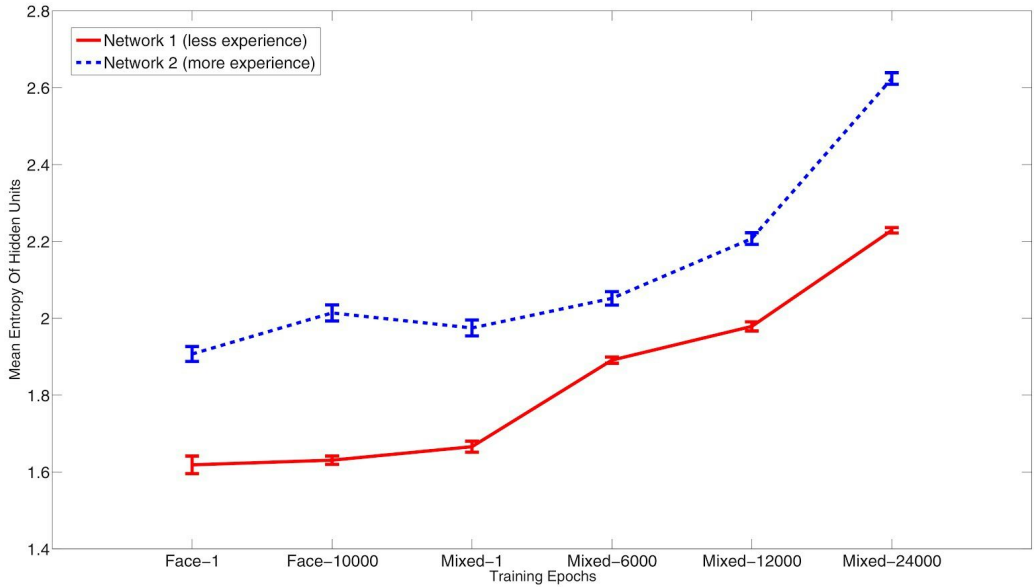

Figure 8. Entropy of hidden units as a function of training epochs. Blue dashed line: network with more experience. Red line: network with less experience. The network with more experience generally has greater entropy across training, suggesting the representation is more separated. Error bars denote ±1 standard error.

## Discussion

Neurocomputational models can provide insight into behavioral results by suggesting hypotheses for how the results came about. We can then analyze the models in ways that are difficult or impossible in human subjects or animals. In this paper, we explored how a neurocomputational model can explain the experience moderation effect observed by Gauthier et al. (2014). We trained networks to perform the same tasks humans have to perform, i.e., to recognize objects and faces. We used one network per subject, setting their parameters based on the individual subject data. We mapped domain general visual ability, $v$, to the number of hidden units, and experience, $E$, to the number of training examples per individual. We showed that the model fits the human data quite well: As the networks gain more experience with the object categories, the correlation between performance on objects and performance on faces increases.

In Experiment 1, as in Gauthier et al. (2014), we had to average across category experience to obtain the correlation with face processing performance. That is, we could not significantly predict face recognition ability based solely on performance within a single category. In Experiment 2, we "recruited" more neural network subjects, and predicted that the effect should hold at the single-category level, provided that there are a sufficient number of subjects that span all levels of visual ability and experience.

Finally, we also attempted to replicate the effect with networks that did not differentiate faces, but simply placed objects and faces into basic-level categories. Here, we did not find an experience-moderation effect, suggesting the type of experience and level of the task (basic or subordinate level discrimination) is an important factor to be considered in understanding these effects.

The conclusion that task matters in terms of the kind of perceptual expertise that is acquired and for the neural substrates recruited is supported by prior work. For instance, novel objects become processed in a holistic manner, like faces, if they are from a category for which subjects practiced individuation, but not categorization (Wong, Palmeri, & Gauthier, 2009a). Likewise, brief individuation training improves discrimination of new faces from another race, while a different training task with the same faces that is as demanding, but does not require individuation, does not improve discrimination (McGugin, Tanaka, Lebrecht, Tarr, & Gauthier, 2011). Qualitatively different patterns of neural representations are observed after training with novel objects in tasks that produce different kinds of behavioral results (Wong et al., 2009b; Wong, Folstein, & Gauthier, 2012).

This experiment predicts that the source of the experience moderation effect is not in regions of the brain that are sensitive only to category level, as opposed to regions that are associated with better performance in individuation for objects and faces, such as the FFA (Furl, Garrido, Dolan, Driver, & Duchaine, 2011; Gauthier et al., 2000; McGugin et al., 2012a; 2014). One advantage of computational models is that we can analyze them in ways we cannot analyze human subjects, to provide hypotheses as to the underlying mechanisms of an effect. For example, an obvious question is, why isn't there a "zero-sum game" between the neurons allocated for each task? I.e., how can the same features be used for both faces, leaves, cars and butterflies?

Behavioral and neural studies show that face recognition and the recognition of other objects of expertise can compete. The N170 face-selective event-related potential is reduced for faces when they are shown in the context of objects of expertise (Gauthier, Curran, Curby, & Collins, 2003; Rossion, Kung, & Tarr, 2004). Behaviorally, non-face objects of expertise compete with faces but not with objects with which subjects are not expert (McKeeff, McGugin, Tong, & Gauthier, 2010; McGugin et al., 2011). fMRI responses to cars in FFA predict behavioral expertise with cars when the cars are presented alone on the screen, and to some degree still when shown among other objects, but not when the other objects are faces (McGugin et al., in press). What all these studies have in common is that interference occurs when faces and objects from another category of expertise have to be processed simultaneously or at least in very close temporal contiguity. Again, this suggests that they are sharing representations.

Our analysis shows why this would be the case. The non-linear transformation by the network at the backpropagation-trained hidden layer displays a spreading transform that separates similar-looking objects. This transform generalizes to new categories. At the same time, as shown in the last four columns in Figure 6, the representation of faces is interdigitated with the representations of other categories. Hence the reason why we see interference in the human subject studies is due to this shared representation. In previous work (Tong et al., 2008), we hypothesized that the FFA contains features useful for fine-level discrimination of faces, and showed how these features generalize to the discrimination of novel categories. Here, we find the same result, shown in the third column of Figure 6, where we find that objects are already separated by the face features, i.e., the transform that separates individual faces also separates individual objects even at the beginning of training on those objects. Given that our model is a model of the FFA, we hypothesize that the location of the experience moderation effect is in the FFA, but more generally, it could be in any area where face representations are more separated.

We conclude that the real power of experience at individuating objects within a homogeneous category is to separate the objects in all dimensions of the representational space spanned by the FFA, and that the experience moderation effect is a direct reflection of this spreading transform. These results support the argument that face and non-face object discrimination are inherently correlated through the sharing of the same mechanism: The better one is at face individuation, the better one will be at individuating objects, given sufficient experience with objects.

One may speculate that one may also find an experience moderation effect at the basic level of categorization. That is, if a subject shows high performance in simply discriminating object categories, and has a great deal of experience in discriminating multiple categories, performance in multiple domains should be correlated. There is some evidence that a great deal of experience with basic level categorization, as in letter recognition, results in a different kind of expertise from that obtained for subordinate-level experience – different both in behavior and neural substrate (Wong & Gauthier, 2007; Wong et al., 2009b). One might hypothesize that multiple $v$'s, i.e., a basic-level $v$ and a fine-level $v$, corresponding to different brain regions associated with these tasks. That is, there must be a constraint that the level of tasks be equalized before one can hope to find such a correlation. In our model, we use fine-level $v$. Evidence for this hypothesis arises in recent work showing that a neural network that is good at differentiating the thousand categories of the Imagenet competition (Russakovsky et al., 2014) develops features that are useful in differentiating other categories (Zeiler & Fergus, 2014; Wang & Cottrell, 2015).

More recently, training backpropagation-based deep neural networks has been shown to achieve state-of-the-art performance on many computer vision tasks, such as image classification (Krizhevsky, Sutskever, & Hinton, 2012; Szegedy et al., 2015), scene recognition (Zhou, Lapedriza, Xiao, Torralba, & Oliva, 2014), and object detection (Girshick, Donahue, Darrell, & Malik, 2014). Researchers also have used deep neural networks to probe representations in neural data, especially in IT (e.g. Cadieu et al., 2014; Yaminis et al., 2014; Güçlü & van Gerven, 2015). Remarkably, these studies have shown that the features learned in the neural networks can explain the representation in human and monkey IT. As these networks are also trained by backpropagation, they support our contention that our neurocomputational model is a reasonable model of FFA and LOC. As a result, it is a promising research direction to use deep neural networks to explain more cognitive/behavioral data, and to model how the brain works.

In summary, we suggest that the correlation between visual processing of faces and objects is mediated by a common representational substrate in the visual system, most likely in the Fusiform Face Area, and that the reason for this mediation is that the FFA embodies a transform that amplifies the differences between homogeneous objects. This transformation is generic; it applies to a wide range of visual categories. The generic nature of this transform explains why there is a synergy between face processing and expert object processing.

## Acknowledgements

This work was supported in part by NSF grant SMA 1041755 to the Temporal Dynamics of Learning Center (TDLC), an NSF Science of Learning Center (GWC and IG), NSF grant IIS-1219252 to GWC, a TDLC trainee grant to PW, and a gift from Hewlett Packard to GWC.